\documentclass[aps,showpacs,twocolumn,nofootinbib]{revtex4-1}
\newcommand{\bb}{\begin{eqnarray}}
\newcommand{\ee}{\end{eqnarray}}
\begin{document}
\title{Area law for black hole entropy in the 
SU(2) quantum geometry approach}
\author{P. Mitra}\email{parthasarathi.mitra@saha.ac.in}
\affiliation{Saha Institute of Nuclear Physics\\ 1/AF Bidhannagar\\
Calcutta 700064}
%\date{1107.4605}
%\widetext
\begin{abstract}
Black hole thermodynamics suggests that a black hole should
have an entropy given by a quarter of the area of its horizon.
Earlier calculations in U(1) loop quantum gravity have led to a
dominant term proportional to the area, but there was a correction
involving the logarithm of the area. We find however that
SU(2) loop quantum gravity
can provide an entropy that is strictly proportional
to the area as expected from black hole thermodynamics.
\end{abstract}
%\bigskip\bigskip\bigskip\bigskip
\pacs{04.60.Pp, 04.70.Dy}
\maketitle
%\narrowtext

\bigskip

A black hole at the classical level is a gravitational field,
a spacetime with a metric satisfying field equations like
those of general relativity. There is a horizon: nothing goes
out of this horizon at the classical level. Quantum considerations
however indicate the emission of particles at a certain
temperature, thereby associating a temperature with the black
hole. For a black hole of mass $M$, this temperature is \cite{hawk}
\bb
T=\frac{1}{8\pi M}.
\ee
Theoretical arguments \cite{bek} have also assigned an entropy to a black
hole, which can be physically understood because the classical
one-way horizon restricts the amount of information available to the 
outside world. This entropy is given by the first law
\bb
T dS=d M
\ee
of thermodynamics to be
\bb
S=4\pi M^2=\frac{A}{4},
\ee
where $A=4\pi (2M)^2$ is the area of the horizon of the black hole.
This relation between the area and the entropy of a black hole
is also valid for rotating and charged cases. 

To understand how the entropy really arises, one would have to
identify quantum states, and before that one would need a
quantum theory of gravity. No full quantum theory is available,
needless to say, but quantum geometry, or loop quantum gravity
has made a good beginning \cite{ash}.

There are gauge fields described by a Chern-Simons theory on the horizon.
One has to visualize cross sections of the horizon having punctures.
Each puncture has an SU(2) spin, with quantum numbers $j,m$ associated.
Geometrical objects like the area become operators in this approach
and have eigenvalues determined by the spin quantum numbers
associated with the punctures. 

The loop quantum gravity approach to black holes has
traditionally fixed the SU(2) gauge group partially to U(1) and 
therefore considered a U(1) Chern-Simons theory \cite{ash}. 
The level $k$ of the Chern-Simons theory is related to the classical horizon 
area, to which the quantum eigenvalue is supposed to be close.
Each puncture contributes to the quantum eigenvalue of the area an amount 
proportional to $\sqrt{j(j+1)}$.  
The spin projection quantum numbers $m$ assigned
to punctures have to satisfy a constraint: the sum of $m$ for
all punctures must be zero. For a given total area, different configurations
of punctures and spins are possible, and these give rise to the number
of states for that area, whence an entropy can be derived. Calculations have
been extensively made in this approach for the number of quantum
states \cite{bhe1,bhe2,bhe3}. 
The entropy is proportional to the area for large area, but
there is a correction term proportional to the logarithm of the area.

Recently it has appeared to be unnecessary to fix the gauge
group, so that the full SU(2) Chern-Simons theory is involved \cite{perez}.
The constraints are different here, because of the SU(2) group. Furthermore,
the level $k$ of the Chern-Simons theory can be kept finite, free and
independent of the area. The aim of the present work was to study the entropy
when the area becomes large and see how far the area law given by 
thermodynamical considerations is obeyed. Calculations can be done
both when the level is made large and when it is kept fixed as the area
is made large. It was discovered that the entropy is strictly proportional
to the horizon area for a fixed level $k$, while the logarithmic
term reappears if the level is made large along with the area. 

In the U(1) formulation, the counting is quite straightforward.
If one considers $p$ punctures on the horizon of a black hole
with $n_j$ punctures carrying spin $j$ for different values of $j$, 
so that $\sum_jn_j=p$, the number of states is
\bb
{(\sum_j n_j)!\over \prod_j n_j!}\prod_j(2j+1)^{n_j}
\ee
without taking the projection constraint into consideration. For
simplicity, let us consider all punctures to have spin 1/2. Then
this becomes $2^p$. Without the projection constraint, the entropy would be
$p\log 2$. But the total projection has to be zero, so $p/2$ spins
have to be {\it up}. The number of ways for this is ${p!\over (p/2)!^2}$,
so the entropy is 
\bb
\log p!-2\log (p/2)!\approx p\log 2-\frac12\log p
\ee
by the Stirling approximation. As each puncture contributes equally
to the area eigenvalue, $p$ is a measure of the area, so that one gets
an area term with a $-\frac12\log$ area correction \cite{bhe1}.

In the SU(2) formulation, the number of states 
for a distribution of spins over punctures 
arises from properties of SU$_q$(2) as
\bb
N=\frac{2}{k+2}{(\sum_j n_j)!\over\prod_j n_j!}
\sum_{a=1}^{k+1}\sin^2 {a\pi\over k+2}\prod_j 
\bigg[{\sin {a\pi(2j+1)\over k+2} \over\sin {a\pi\over k+2}}\bigg]^{n_j}
\label{long}\ee
\cite{log,perez}\footnote{The
combinatorial factor arises from the choice of the punctures.}. 
If we take $j_i=1/2$ for each puncture,
\bb
N=\frac{2}{k+2}\sum_a\sin^2 {a\pi\over k+2} [2\cos {a\pi\over k+2}]^p.
\label{half}\ee
It is of interest to see what happens if the level $k$
is made large. In this case, the number of terms in the sum over $a$
goes to infinity and it is possible to approximate the sum by an integral.
One obtains
\bb
N=\frac{2}{k+2}\int_0^k da \sin^2 {a\pi\over k}2^p\cos^p {a\pi\over k}.
\ee
For odd $p$, the integral vanishes, while
for even $p$, the range of integration can be halved:
\bb
N=\frac{4}{\pi}\int_0^{\pi/2} dx\ \sin^2 x\ 2^p\ \cos^p x.
\ee
This involves a well known integral:
\bb
N=\frac{4}{\pi}2^p\frac{\pi}{2}{(p-1)!!\over (p+2)!!}.
\ee
Now the Stirling approximation for large $n$,
\bb
n!\approx \sqrt{2\pi n}n^ne^{-n},
\ee
yields  
\bb
N\approx 2^{p+3/2}p^{-3/2}\pi^{-1/2}.
\ee
This means the entropy is
\bb
\log N\approx p\log 2-\frac32\log p.
\ee
Thus there is an area term and a logarithmic correction with coefficient -3/2
\cite{log, carlip, perez}.

Next we consider $k$ to be fixed in (\ref{half}) when $p$ is made large.
The argument of the sine/cosine varies from term to term, but the finite sum is 
dominated by the largest term, which occurs for the largest value of 
$|\cos{a\pi\over k+2}|$ in the available range of $a$. 
The number of punctures $p$ occurs only in the exponent in $N$: 
\bb
N=\frac{4}{k+2}\sin^2 {\pi\over k+2} [2\cos {\pi\over k+2}]^p.
\ee
But the area of the horizon is proportional to $p$, 
hence the entropy 
\bb
\log N\propto {\rm area}. 
\ee
To see whether the terms with other values of $a$ vitiate this result, one just
has to note that for $0<y<x$
\bb
\log (x^p+cy^p)\approx p\log x+c\exp (-p\log\frac{x}{y}),
\label{dom}\ee
so that the corrections are (a finite number of) exponentially
small terms with no $\log p$.
This means that there is no logarithmic correction  of the
usual type when all 
spins are taken to be 1/2 and $k$ is held constant as $p\to\infty$.

These arguments can be extended to a general distribution of spins.
In this case, we write (\ref{long}) as
\bb
N&=&\frac{2}{k+2}{(\sum_j n_j)!\over\prod_j n_j!}
\sum_{a=1}^{k+1}\sin^2 {a\pi\over k+2}\prod_j [f_j]^{n_j},\nonumber\\
&=&\frac{2}{k+2}{(\sum_j n_j)!\over\prod_j n_j!}
\sum_a\sin^2 {a\pi\over k+2} F(\cos {a\pi\over k+2}),
\label{full}\ee
where 
\bb
f_j(\cos {a\pi\over k+2})\equiv
{\sin {a\pi(2j+1)\over k+2} \over\sin {a\pi\over k+2}}
\ee
and
\bb
F(\cos {a\pi\over k+2})\equiv\prod_j [f_j]^{n_j}.
\label{F}\ee

Let us first consider $k$ becoming large, so that the sum over $a$ can
be treated as an integral. As $a$ is varied, the integrand
\bb
\sin^2 {a\pi\over k+2} [F(\cos {a\pi\over k+2})]
\ee
attains its maximum when
\bb
2\cot {a\pi\over k+2}=\sin {a\pi\over k+2}{F'\over F}.
\ee
At this maximum, $a$ satisfies
\bb
({a\pi\over k+2})^2\approx 2{F(1)\over F'(1)},
\ee
which is small because $F'$ contains the numbers $n_j$ which have to be large 
for large area.
As ${a\pi\over k+2}$ is small, the integrand can be approximated as
\bb
({a\pi\over k+2})^2[F(1-\frac12({a\pi\over k+2})^2)].
\ee
However, to get an idea of the integral itself, one needs to know
the width of the peak, which can be estimated by evaluating the
second derivative of this expression. The result is
\bb
\frac{2\pi^2}{(k+2)^2}F-\frac{5\pi^4a^2}{(k+2)^4}F'
+\frac{\pi^6a^4}{(k+2)^6}F'',
\ee
which, for large $n_j$, simplifies at the maximum to
\bb
-4\frac{\pi^2}{(k+2)^2}F.
\ee
Consequently, the width $\sigma$ of the peak is given by
\bb
\sigma^2=
\frac{(k+2)^2}{\pi^2}\frac{F}{F'},
\ee
which is essentially the ratio of the maximum to the second derivative
and the integral can be approximated by
\bb
\frac{2(k+2)}{\sqrt\pi}(\frac{F}{F'})^{3/2}F.
\ee
Now one has to take the 
combinatorial factor of the punctures into account. 
The number of states for the distribution can be written as
\bb
N=\frac{4}{\sqrt\pi}{(\sum_j n_j)!\over\prod_j n_j!}
(\sum_jn_j\frac{f_j'}{f_j})^{-3/2}\prod_j [f_j]^{n_j}
\ee
by expanding $F$ in terms of the $f_j$.
To maximize this number, one sets 
\bb
\delta\log N=0
\ee
when the numbers 
$n_j$ of punctures with
spin $j$ are varied, subject to the constraint of fixed area
\bb
\sum_j\sqrt{j(j+1)}\delta n_j=0.
\ee
One obtains for large $n_j$
\bb
\log f_j+\log \sum_k n_k-\log n_j-\lambda\sqrt{j(j+1)}=0,
\ee
where $\lambda$ is a Lagrange multiplier. One finds then
\bb
n_j=(\sum_k n_k)f_j e^{-\lambda\sqrt{j(j+1)}},
\ee
whence, for consistency,
\bb
\sum_{j=1/2}^{k/2} f_j e^{-\lambda\sqrt{j(j+1)}}=1.
\label{lambda}\ee
It has to be emphasized that $j$ goes from $\frac12$ to $\frac{k}{2}$ for
a level $k$, so that for $k\to\infty$ the upper limit is infinite.
In this case, furthermore, $f_j$ reduces to $2j+1$ as $\frac{a}{k+2}\to 0$
for large area. Now the area 
\bb
A=\sum_jn_j\sqrt{j(j+1)}
\ee
in appropriate units. 
Taking the above distribution, one easily sees that
\bb
\log N=\lambda A-\frac32\log A \quad {\rm for~}k\to\infty
\ee
as each $n_j$ goes like the area for large area. 
Constant terms are neglected here.

Next, we suppose $k$ to be kept fixed and the area made large. The sum
(\ref{full}) over the finite number of values of $a$
can be considered term by term. For each $a$,
the maximization of $N$ with respect to the $n_j$ goes through as above.
One gets
\bb
\log N=\lambda A \quad {\rm for~finite~} k
\ee
where $\lambda$ is still determined by (\ref{lambda}) with $f_j$ evaluated
for the $a$ under consideration and the sum restricted to $j\leq k/2$. 
Now $\lambda$ depends on $k$ \cite{perez} and also on $a$. 
When the summation of $N$ over $a$ is envisaged, the logic of (\ref{dom})
shows that the highest $\lambda$ dominates and
the contributions of values of $a$ giving lower $\lambda$ are exponentially
suppressed for large area.
In other words, $\lambda$ has to be maximized over $a$, which determines
the relevant value(s) of $a$.
There is no logarithmic correction in the entropy because
the $(\frac{F}{F'})^{3/2}$ factor, which appeared with $a$ taken to be
continuous, does not appear in this case of finite $k$ and discrete $a$.

These calculations of the entropy
are for the most probable distribution, but the sum over all
distributions can also be estimated. There are correction factors
proportional to the area from the factorials and from the integrations
approximating the sums over $n_j$, but these actually cancel out
because there is only one constraint, {\it viz.}, the area constraint
(cf. \cite{bhe2}).

In conclusion, if instead of making the level $k$ grow, 
one holds it fixed and allows the black hole horizon area 
to become large, the black hole entropy is strictly proportional
to the area without a logarithmic correction.
If on the contrary the level $k$ is allowed to go to infinity,
the usual SU(2) logarithmic correction with a coefficient -3/2 is recovered,
in accordance with \cite{log, perez}.
What is new is that the possibility, suggested by \cite{perez}, 
of keeping the level $k$ of the Chern-Simons theory on the horizon fixed, 
yields a behaviour of the 
entropy which is seen to satisfy the thermodynamic area law accurately.
The earlier calculations \cite{log, perez} mostly considered $k$ going to
infinity, and obtained the log correction which we have also reproduced
in that limit. Fixed $k$ was considered in \cite{perez} using Laplace
transforms and assumptions of analytic continuation, the {\it same} logarithmic
correction being obtained as for $k$ going to infinity, but they noted that
their arguments were less than rigorous.
We have explained above why there has to be a difference in the two limits:
the factor of $({F\over F'})^{3/2}$, which is responsible for the logarithmic
correction, occurs only for $k\to\infty$.

Both behaviours are in contrast to the distinctive coefficient -1/2 for the
logarithmic correction in the U(1) theory.
It may be recalled that in the U(1) counting, there is an extra constraint,
the sum over spin projections at all punctures being zero, and that
produces a $-\frac12\log A$ correction. This is seen explicitly in the
case when all spins are set equal to $\frac12$. More generally, when
the number of states is maximized for a distribution of spins and the
distribution summed over, the fact that there are two constraints, the area
constraint and the spin projection constraint, produces the 
$-\frac12\log A$ correction \cite{bhe2}. 
This summation does not produce any correction
in the SU(2) case, as pointed out above, because there are no more constraints
to be taken care of apart from the area constraint. The SU(2) constraints
are already incorporated into the sum over $a$ in (\ref{long}). The SU(2) 
correction arises from two sources: one is the $\sin^2$ factor in (\ref{long}),
and the other is the width of the peak in the sum over $a$. Both of these
come into play only in the limit $k\to\infty$ when the sum is converted to
an integral. The $\sin^2$ factor is responsible for a $1/A$ suppression in $N$
because $\frac{a}{k+2}$ becomes small for large $k$. The
width of the peak decreases like $1/\sqrt{A}$ as the area increases
and $k\to\infty$. For finite $k$, the summation collapses to one or two terms
and neither suppression factor
appears, so that there is no logarithmic correction.

It must be noted that the U(1) calculation would produce the logarithmic
correction even if $k$ could be kept finite, because the correction comes
from a different source there, namely from the spin projection constraint.
Thus the possibility of having a strict area law exists only for the
SU(2) theory.

The level $k$ is normally a measure 
of the classical area of the horizon, while $A$ is the eigenvalue of the
area operator. The terms in the entropy therefore relate to the
quantum eigenvalue of the area. The logarithmic correction, which 
arises when $k$ becomes large, might be thought to be
caused by a large classical area, being absent when the level
$k$ is kept finite and the quantum eigenvalue allowed to soar. However,
this limit does not require the classical area to be kept fixed, as quantization
ambiguities were exploited \cite{perez} to delink the level $k$ from the
classical area. One can thus {\it choose} to quantize SU(2) loop
quantum gravity in such a way that the entropy grows linearly with the
area exactly as predicted by thermodynamical ideas.
This is not too dissimilar from the choice made in most versions of loop quantum
gravity when the Immirzi parameter \cite{ash} is fixed to make the 
leading contribution to the entropy
a quarter of the horizon area instead of being merely proportional to it.

In other words, loop quantum gravity can be tailored to
yield the precise area dependence of the entropy predicted from
thermodynamical considerations without any logarithmic correction.
This is important because no correction has been found to
the Hawking temperature \cite{tunnelcorr} which is at the root of the
thermodynamical area law.
\bigskip

I thank Amit Ghosh for discussions.

\end{document}